\documentclass[useAMS]{mn2e}
\usepackage{graphics,mncite}
\usepackage{times}

\newcommand{\ms}{m\,s$^{-1}$} 
   
\newcommand{\equ}{$\gamma$~Equ}
\newcommand{\cir}{$\alpha$~Cir}

\newcommand{\ion}[2]{#1\,{\sc #2}}
\newcommand{\lnd}{$\overline{g}$}
\newcommand{\lan}{\ion{La}{ii}}
\newcommand{\pr}{\ion{Pr}{iii}}  
\newcommand{\nd}{\ion{Nd}{iii}}
\newcommand{\eu}{\ion{Eu}{ii}}
  
\newcommand{\feo}{\ion{Fe}{i}} 
\newcommand{\fet}{\ion{Fe}{ii}}

\newcommand{\bz}{$B_\ell$}
\newcommand{\bs}{$B_s$}
\newcommand{\ub}{$^{\bf b}$}
\newcommand{\ur}{$^{\bf r}$}
\newcommand{\uc}{$^{\bf c}$}

\newcommand{\fifps}[2]{\centering\resizebox{#1}{!}{\includegraphics{#2}}}

\begin{document}

\title[No pulsational variations of the surface magnetic field in $\gamma$~Equulei]
{The null result of a search for pulsational variations of the surface magnetic field in the roAp star $\gamma$~Equulei
}

\author
[O. Kochukhov et al.]
{O. Kochukhov$^1$\thanks{E-mail: kochukhov@astro.univie.ac.at},
T. Ryabchikova$^{2,1}$, J. D. Landstreet$^3$, W. W. Weiss$^1$ \\
$^1$Department of Astronomy, University of Vienna, T\"urkenschanzstra{\ss}e 17, 1180 Vienna, Austria \\
$^2$Institute of Astronomy, Russian Academy of Sciences, Pyatnitskaya 48, 119017 Moscow, Russia \\
$^3$Department of Physics and Astronomy, University of Western Ontario, London, Ontario N6A 3K7, Canada \\
} 

\date{Accepted 0000 January 00. Received 0000 January 00; in original form 0000 January 0}

\maketitle
\pubyear{2004}

\begin{abstract} 
We describe an analysis of the time-resolved measurements of the surface
magnetic field in the roAp star \equ. We have obtained a 
high-resolution and high S/N spectroscopic time-series, and the magnetic field was
determined using Zeeman resolved profiles of the \fet\ 6149.25~\AA\ and 
\feo\ 6173.34~\AA\ lines. Contrary to recent reports we do not find any
evidence of magnetic variability with pulsation phase, and derive an upper limit
of 5--10~G for pulsational modulation of the surface magnetic field in \equ.
\end{abstract}


\begin{figure*}
\fifps{15.4cm}{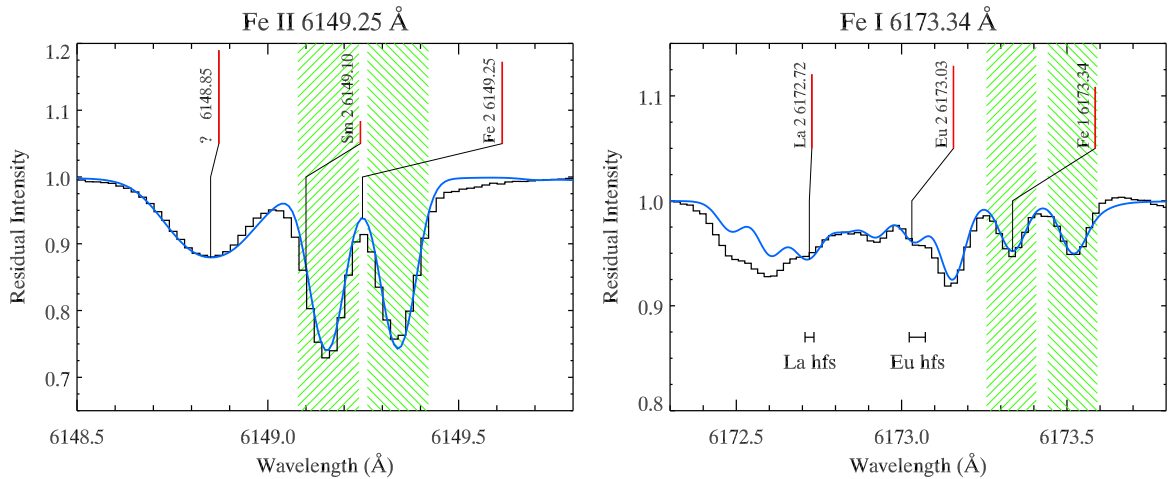}
\caption{Comparison of the average CFHT spectra of \equ\ (histogram) in
the 6149 (left panel) and 6173~\AA\ (right panel) regions with magnetic spectrum
synthesis (solid line). The length of the line underlining each identification is
proportional to the strength of corresponding spectral feature. The
\ion{La}{ii} 6172.72~\AA\ and \ion{Eu}{ii} 6173.03~\AA\ lines were computed
taking into account hyperfine splitting. For clarity we do not show individual
hfs components, but indicate the range of their central wavelengths with
horizontal bars. The shaded areas illustrate our selection of the wavelength
intervals for the centroid line position measurements.} 
\label{fig1}
\end{figure*}

\begin{table*}
\begin{minipage}{160mm}
\caption{Journal of spectroscopic observations of \equ.\label{tbl1}}
\begin{tabular}{ccccccccc}
\hline
UT date & Instrument/& Wavelength   & Resolution               & Exposure & Number of & Start HJD & End HJD   & Typical \\
        & telescope  & region (\AA) & ($\lambda/\Delta\lambda$)& time (s) & exposures & (2450000+)& (2450000+)& SNR     \\
\hline
22/07/1999 & CES/ESO 3.6-m & 6140--6165 & 166\,000 & 60 & 31 & 1381.78344 & 1381.82390 & 190 \\
26/09/2002 & Gecko/CFHT    & 6104--6194 & 115\,000 & 90 & 64 & 2543.82314 & 2543.92191 & 230 \\
\hline
\end{tabular}
\end{minipage}
\end{table*}

\section{Introduction}
\label{introduction}

After discovery of the conspicuous radial velocity (RV) pulsational variations
in a sample of rapidly oscillating magnetic peculiar (roAp) stars (Kanaan \&
Hatzes \scite{KH98}, Savanov, Malanushenko \& Ryabchikova \scite{SMR99}, 
Kochukhov \& Ryabchikova \scite{KR01a} for \equ; Baldry et al. \scite{BBV98},
Baldry \& Bedding \scite{BB00}, Kochukhov \& Ryabchikova \scite{KR01b} for \cir\
and HD~83368), attempts to search for magnetic field variations over the
pulsational period have been made. First, Hubrig et al. \cite{HKB04}  tried to
measure pulsational variability of the longitudinal magnetic field \bz\ in six
roAp stars. Their sample included \equ\ -- probably one of the most favourable
stars for this kind of investigation. \equ\ is a bright northern roAp star
with a strong magnetic field, and with one of the largest pulsational RV
amplitudes which exceeds 1000~\ms\ in individual spectral lines. The extremely slow rotation of \equ, leading to 
very sharp spectral lines, makes this star the best candidate for any study of
the pulsational variability in spectroscopy. Hubrig et al. \cite{HKB04} used
low-resolution Zeeman time-series observations and measured \bz\  using
hydrogen lines and unresolved blends of metal lines. They failed to detect any
variability beyond the formal errors of their measurements which were
40--100~G. 

According to a coarse theoretical estimate made by Hubrig et al. \cite{HKB04}
there should exist a linear relation between magnetic field variability over 
the pulsational cycle and the RV amplitudes.  In roAp stars the largest RV amplitudes
are observed in the lines of first and second ions of rare-earth elements
(REE), while they are usually below measurement errors for the lines of iron
group elements. Thus, high-resolution spectroscopy and spectropolarimetry of
selected REE spectral lines is a more promising tool for an investigation of
possible rapid magnetic oscillations in roAp stars. Taking this into account
Leone \& Kurtz \cite{LK03} obtained a high-resolution (R=115\,000), high S/N
time-series of observations of \equ\ with a circular polarization analyzer, and
measured \bz\ using four \nd\ lines. They reported the discovery of pulsational
variations with amplitudes between 112 to 240~G and, more surprisingly,
with discrepant phases of magnetic maximum for different \nd\ lines. Leone \&
Kurtz's result was based on only 18 time-resolved spectra.  A year later
Kochukhov, Ryabchikova \&  Piskunov \cite{KRP04} obtained a time-series of
polarimetric observations of \equ\ with a smaller resolving power (R=38\,000),
but acquired more than 200 spectra over 3 nights, more than compensating for
lesser quality of individual spectra. Kochukhov et al. \cite{KRP04} used
simultaneously 13 \nd\ lines for magnetic measurements which allowed them to
achieve a substantial reduction of the error of the \bz\ determinations. They did
not confirm longitudinal field variability over the pulsational period in \equ\
and gave a conservative upper limit of $\approx$40~G for the amplitude of
pulsational modulation of \bz\  determined from \nd\ lines. 

Another attempt to search for possible rapid magnetic variability in \equ\ was
made by Savanov, Musaev \& Bondar \cite{SMB03}. They measured the surface
magnetic field \bs\ variations over the pulsational period using the \fet\
6149.25~\AA\ line observed in unpolarized light. Due to a very simple Zeeman
pattern (two equally separated $\pi$- and $\sigma$-components) this line is
ideal for \bs\ measurements (see Mathys et al. \scite{MHL97}). Savanov et al.
reported a 1.8$\sigma$ detection of \bs\ variability with an amplitude of
99$\pm$53~G. At the same time they did not find periodic variations of RV
measured for the individual Zeeman resolved components of the \fet\
6149.25~\AA\ doublet exceeding their error limit (100--120~\ms). The
authors used high-resolution R=120\,000 time-series observations, but the S/N of a
single spectrum did not exceed 40--60. Clearly, the result of Savanov et
al.~\cite{SMB03} is marginal and needs to be confirmed or rejected with data of
better quality.   

In this paper we present the results of a new search for pulsational variations
of \bs\ in \equ\ using high-resolution and high S/N time-resolved observations
of this star. These observational data allowed us to obtain precise measurements
of the magnetic field in \equ, and strongly constrain possible changes of \bs\ 
over the pulsation cycle of the star.  

\begin{figure*}
\fifps{14.5cm}{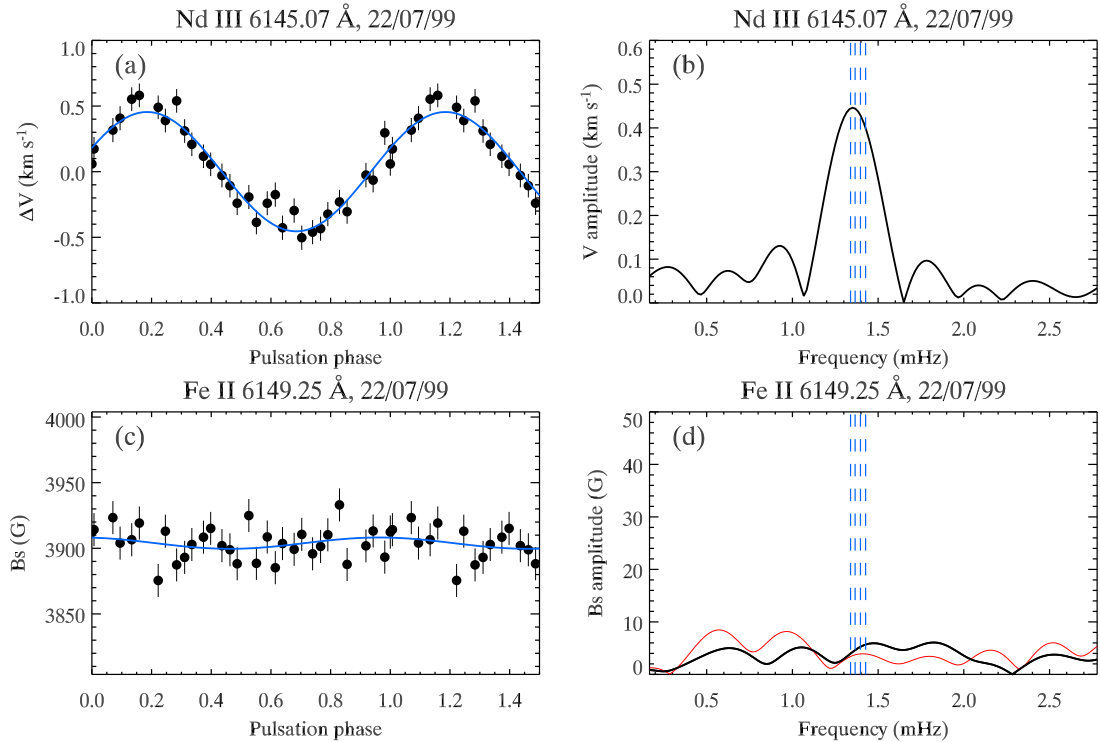}
\caption{Measurements of radial velocity and surface
magnetic field obtained for \equ\ on 22/07/99. The \nd\ 6145.07~\AA\ 
pulsational RV curve (a) is folded with the best-fit
oscillation period and is compared (c) with the variation of \bs\ 
measured from the separation of the Zeeman components of the \fet\ 
6149.25~\AA\ (centre-of-gravity line position measurements). Panels (b) and (d) illustrate the amplitude 
spectra for the RV and \bs\ (thick line -- centre-of-gravity measurements,
thin line -- multiple fit of gaussians). The vertical dashed lines show the photometric 
pulsational periods of \equ\ (Martinez et al. 1996).}
\label{fig2}
\end{figure*}

\section{Observations and data reduction}
\label{observations}

The time-resolved observations of \equ\ were obtained using the single-order
$f/4$ Gecko coud\'e spectrograph with the EEV1 CCD at the 3.6-m
Canada-France-Hawaii telescope. Table~\ref{tbl1} gives details of this
spectroscopic time-series dataset. The spectra cover approximately the
spectral window 6104--6194~\AA. This wavelength interval contains two Zeeman
resolved lines, \fet\ 6149.25~\AA\ and \feo\ 6173.34~\AA. It also has strong \nd\ and \pr\
lines, optimal for investigation of pulsational variability of roAp stars, as
well as strong and weak lines of other elements such as Si, Ca, Cr, and Ba.

The spectra were reduced using standard IRAF tasks. Each stellar, flat and
calibration frame had a mean bias subtracted and was then cleaned of cosmic ray
hits and extracted to one dimension. Extracted stellar spectra were divided by
an extracted mean flat field, and the continuum was fit with a third-order Legendre 
polynomial, using the same rejection parameters for all spectra so that the
continuum fit is as uniform as possible. The wavelength scale was established
using about 40 lines of a ThAr emission lamp, resulting in an RMS scatter about
the adopted pixel-wavelength polynomial (a sixth-order Legendre polynomial) of about $5~\times~10^{-4}$ \AA. The
wavelength scale was linearly interpolated between ThAr lamp spectra taken
before and after the stellar series, but the spectra were not resampled to a 
linear wavelength spacing. 

In this paper we also used 31 very high resolution time-resolved observations
of \equ\ analysed by Kochukhov \& Ryabchikova \cite{KR01a}. These time-resolved
data were obtained with the Coud\'e Echelle Spectrograph (CES), fiber-linked to
the Cassegrain focus of the ESO 3.6-m telescope. The highest resolution CES
image slicer and the ESO CCD\#38 were used, allowing us to reach a resolving power
of  $\lambda/\Delta\lambda=166\,000$ and record spectra in the 6140--6165~\AA\
wavelength interval. We refer the reader to Kochukhov \& Ryabchikova
\cite{KR01a} for other details of the acquisition and reduction of the CES
spectra of \equ. 


\begin{table}
\caption{Results of the analysis of RV pulsational 
variability of \equ. The table gives pulsation period $P$, amplitude 
$\Delta RV$ and phase $\varphi$
(measured in units of oscillation period). Superscripts ``b'', ``r''
and ``c'' denote respectively the blue-shifted, red-shifted and central
components of the Zeeman resolved Fe lines. \label{tbl2}}
\begin{tabular}{llcc}
\hline
~~~~~Line      &  ~~$P$ (min)  & $\Delta RV$ (\ms) & $\varphi$ \\
\hline
\multicolumn{4}{c}{\it 22/07/99, CES/ESO 3.6-m}\\
\nd\  6145.07    & 12.290$\pm$0.071 & 454.6$\pm$23.4 & 0.815$\pm$0.016 \\
\fet\ 6149.25\ub & 12.290           &  10.1$\pm$5.7  & 0.383$\pm$0.089 \\
\fet\ 6149.25\ur & 12.290           &   9.4$\pm$5.1  & 0.215$\pm$0.085 \\
\multicolumn{4}{c}{\it 26/09/02, Gecko/CFHT}\\
\nd\  6145.07    & 12.281$\pm$0.039 & 165.1$\pm$10.8 & 0.462$\pm$0.021 \\
\fet\ 6149.25\ub & 12.281           &   3.6$\pm$2.5  & 0.000$\pm$0.113 \\
\fet\ 6149.25\ur & 12.281           &   5.2$\pm$3.2  & 0.415$\pm$0.219 \\
\feo\ 6173.34\uc & 12.281           &   5.4$\pm$14.9 & 0.222$\pm$0.143 \\
\feo\ 6173.34\ur & 12.281           &  17.2$\pm$15.8 & 0.628$\pm$0.234 \\
\hline
\end{tabular}
\end{table}

\section{Magnetic field and radial velocity measurements}

As a first step in our analysis of \equ\ we computed synthetic spectra in the
region of the Zeeman resolved iron lines \fet\ 6149.25~\AA\ and \feo\
6173.34~\AA\  (hereafter we refer to these spectral features simply as the
\fet\ and \feo\ lines) using abundances, Fe stratification and model atmosphere 
from the study of Ryabchikova et al. \cite{RPK02}.  
Synthetic spectra of \equ\ were calculated with the \mbox{\sl SYNTHMAG} code 
(Piskunov~\scite{PISK99}), modified to take into account vertical 
stratification of chemical abundances. 
The simplest model, with a constant magnetic field over the stellar surface, 
was adopted. The splitting pattern of the \feo\ line is a pure Zeeman triplet with unshifted central
$\pi$-component and two $\sigma$-components.
To fit the relative intensity of the $\pi$- and $\sigma$-components of this line
the magnetic field vectors in our model have to be inclined 
by $\approx50\degr$ to the stellar surface. 
To look more carefully at the blending effects we took into account
the hyperfine structure of nearby \lan\ 6172.72~\AA\ and \eu\ 6173.03~\AA.
An unidentified line at $\lambda$~6148.85~\AA, which shows a pulsational behaviour 
typical for REE lines, was synthesized with arbitrary atomic parameters 
to take into account its possible blending of the blue-shifted \fet\ component.

The comparison of these calculations with the average CFHT spectrum
is presented in Fig.~\ref{fig1}. This figure shows that the \fet\ line is free from
significant blends, which enables accurate pulsational analysis  of both the
red and blue  components of this Zeeman doublet. 
In particular, we found that pulsational variability of
the 6148.85~\AA\ feature has negligible influence on the measurements of the
blue component of the \fet\ line. On the other hand, the blue
$\sigma$-component of the \feo\ line is blended by the \ion{Eu}{ii}
6173.03~\AA\ and hence is not suitable for time-resolved measurements of \bs.
Consequently, RV analysis and field modulus measurements using the \feo\ line
were based on the $\pi$- and red $\sigma$-component. 

\begin{figure*}
\fifps{14.5cm}{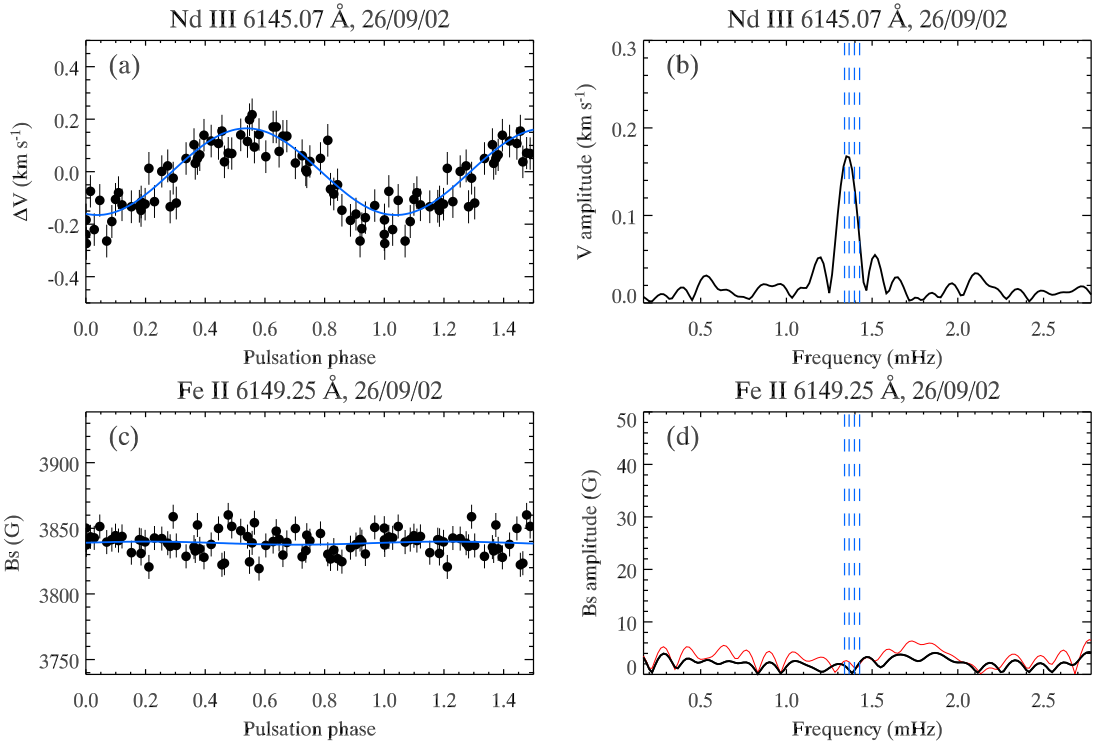}
\caption{The same as Fig.~\ref{fig2} for the RV and \bs\ determined with 
the time-resolved spectra of \equ\ obtained on 26/09/02.}
\label{fig3}
\end{figure*}

\begin{figure*}
\fifps{14.5cm}{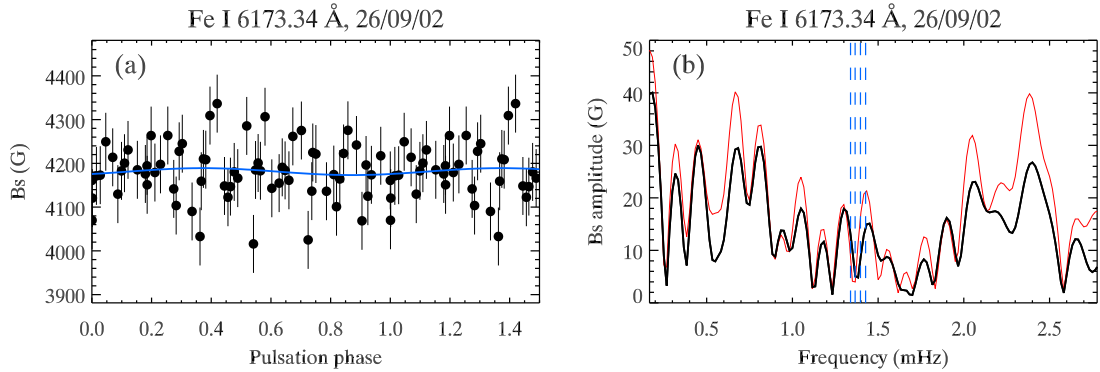}
\caption{Magnetic field measurements obtained for \equ\ on 26/09/02
using the \feo\ 6173.34~\AA\ line. The panel (a) shows \bs\ 
phased with the best-fit pulsation period, the panel (b) presents the 
amplitude spectrum for \bs\ (thick line -- centre-of-gravity measurements,
thin line -- multiple fit of gaussians).}
\label{fig4}
\end{figure*}

We determined the time-dependent position of the centres of Zeeman components using
centre-of-gravity measurements within the spectral regions indicated in Fig.~\ref{fig1}. 
We note that an alternative
technique of fitting a superposition of three gaussian profiles 
to the 6148.85~\AA\ and \fet\ line represents a better way
to derive \bs\ from \textit{time-averaged} spectra because it allows to model partially resolved
\fet\ components and remove the blending contribution of the 6148.85~\AA\ feature. 
However, this method may encounter difficulties in describing \textit{time-resolved} 
profiles of REE lines, including the 6148.85~\AA\ line, which show substantial
RV shifts and exhibit profile asymmetries associated with non-radial pulsation velocity field.
Consequently, we prefer to use the centre-of-gravity technique throughout this paper
and show the gaussian fitting results in Figs.~\ref{fig2}--\ref{fig4} for comparison
purpose only.

The RV obtained from the iron lines were compared with the outstanding
pulsational variation seen in the \nd\ 6145.07~\AA\ line. 
This feature allows us to verify the presence of rapid spectroscopic
variability during our observations of \equ\ and to determine the oscillation 
period and amplitude with high accuracy. For each available
dataset we established the period of the spectroscopic pulsational variation
using this \nd\ line and then fitted the phase (counted from the start HJD,
see  Table~\ref{tbl1}) and the amplitude of the Fe RV measurements using a
non-linear least-squares technique.  Results of this analysis are summarized in
Table~\ref{tbl2}. We do not find any pulsational variation of the RV determined
for the Zeeman resolved components of the Fe lines exceeding about 5--10~\ms\
for the \fet\ and $\approx20$~\ms\ for the \feo\ line. 

The measurements of the \fet\ line reported in Table~\ref{tbl2} supersede tentative RV
amplitude of 64~\ms\ given by Kochukhov \& Ryabchikova \cite{KR01a}, who used the same
CES dataset as studied here. This difference mainly comes from the fact that in the
present paper we adopt a fixed pulsation period ($P=12.29$~min) for the analysis of the
individual \fet\ Zeeman components, while a set of  four photometric periods was tested
in our previous paper and the corresponding  maximum RV amplitude for $P=11.93$~min
and for the whole line was reported.

The measurements of the line centre positions were converted to surface field
using the expression:
\begin{equation}
B_s = \frac{\Delta\lambda_{\rm mag}}{4.67\times10^{-13} \lambda_0^2 \overline{g}},
\label{eqn1}
\end{equation}
where $\Delta\lambda_{\rm mag}$ is the half of the wavelength difference between the red
and blue Zeeman components of the \fet\ line or the distance between the
red $\sigma$- and the $\pi$-component of the \feo\ line, \lnd\ is the
mean Land\'e factor and $\lambda_0$ is the laboratory wavelength of a line. 
Based on the information available from the VALD database (Kupka et al. \scite{KPR99}) 
we adopted
$\overline{g}=2.5$ and 1.35 for the \feo\ and \fet\ lines respectively. The internal
precision of our magnetic measurements is estimated to be $\sim$10~G for the \fet\
line and $\approx$70~G for the weaker \feo\ line. At the same time the average
field strengths derived from the two lines are substantially different in the CFHT
spectra: centre-of-gravity measurements of the \fet\ line give 
$3839\pm9$~G ($3935\pm14$~G is obtained with multiple fit of three gaussians), while 
$\langle B_s \rangle = 4181\pm66$~G ($4200\pm84$~G) is derived using the weaker \feo\ line.  
This $\approx300$~G difference is confirmed by the spectrum synthesis. 
The discrepant \bs\ might be related to inaccuracy of the tabulated Land\'e
factors, or it could be real and reflect different horizontal and/or vertical formation
regions of the two diagnostic lines, or the effects of saturation in the \fet\ line. 
An average field strength of $3903\pm13$~G ($3941\pm19$~G) was
obtained from the \fet\ line in the ESO spectra of \equ.

\begin{table}
\caption{Results of the analysis of time-resolved \bs\ measurements for
\equ. The columns give spectral lines used for \bs\ determination,
magnetic pulsation amplitude $\Delta$\bs, phase $\varphi$ and RMS scatter
$\sigma$\bs\ of \bs. The last column reports an  amplitude $\delta$\bs\ of
surface field variability which would have been detected with our data at the
3$\sigma$ confidence level. \label{tbl3}}
\begin{tabular}{llccc}
\hline
~~~~~Line     & ~$\Delta$\bs\ (G) & $\varphi$ & $\sigma$\bs\ (G)& $\delta$\bs\ (G)\\
\hline
\multicolumn{5}{c}{\it 22/07/99, CES/ESO 3.6-m}\\
\fet\ 6149.25 & 4.3$\pm$3.2  & 0.037$\pm$0.120 &  13.0  & 13 \\
\multicolumn{5}{c}{\it 26/09/02, Gecko/CFHT}\\
\fet\ 6149.25 & 1.2$\pm$1.6  & 0.294$\pm$0.219 &   9.2  & 5  \\
\feo\ 6173.34 & 7.9$\pm$11.8 & 0.628$\pm$0.234 &   66.4 & 28 \\
\hline
\end{tabular}
\end{table}

Table~\ref{tbl3} and Figs.~\ref{fig2}--\ref{fig4} present results of our
time-series analysis of the surface field measurements. We see absolutely no
evidence for any magnetic variations during the pulsation cycle in \equ. Formal results
of the least-squares fits with a fixed pulsation period as derived from the \nd\
line indicate insignificant amplitudes, all below 10~G.  Our most precise
time-resolved  magnetic measurements are derived from the CFHT dataset. Analysis of
the 6149.25~\AA\ line in these spectra results in formal \bs\ amplitude of just
$\approx$1~G, with the highest noise peaks in the amplitude spectrum (Fig.~\ref{fig3}d)
not exceeding 5~G over the whole period domain typical for roAp pulsations.

A 3$\sigma$ upper limit of the amplitude of magnetic variability was estimated
from Monte Carlo  simulations by sampling a sinusoidal signal at the phases of
our  observations and adding a random  noise, characteristic of the scatter of
the \bs\ measurements. This rigorous statistical estimate is reported in
Table~\ref{tbl3} and indicates that, at the 3$\sigma$ confidence level, no
magnetic variability with the amplitude $\ge$5~G is seen in \equ\ during the night
of our CFHT observation.

\section{Conclusions}

Our time-resolved magnetic measurements of \equ\ have achieved the highest
precision for a roAp star, but reveal no evidence of pulsational modulation of
the field strength. We constrain possible magnetic variability to be below
$\approx$5~G in Fe lines. These results suggest that the marginal detection of
$\approx$100~G \bs\ variability during the pulsation cycle  reported in \equ\ by
Savanov et al. \cite{SMB03} is spurious, and probably stems from insufficient
precision of the \bs\ measurements in that study.

The null result reported in the present paper complements non-detection of the
pulsational  variability of \bz\ determined from \nd\ lines (Kochukhov et al.
\scite{KRP04}). It should be recalled that the Fe lines studied in our
paper and the REE lines showing strong pulsational RV modulation are formed at
substantially different atmospheric depths due to the extreme
stratification of chemical abundances in cool Ap stars and in \equ\ in particular.
Stratification analysis presented by Ryabchikova et al. \cite{RPK02} allows us
to conclude that \nd\ lines sample very high atmospheric layers with optical
depths $\log\tau_{5000}\la-7$, while the Fe  lines are formed below
$\log\tau_{5000}\approx-1$. The striking difference in the RV amplitudes of Fe and
\nd\ lines is then attributed to an outward increase of pulsational amplitude
by roughly a factor of 50--100. This increase is none the less not
accompanied by an increase or even the presence of any observable oscillations of
the magnetic field structure. Combining the results of this paper with the study of 
Kochukhov et al. \cite{KRP04}, and taking into account that for \equ\
$B_s\approx  2.6 B_\ell$, we find that at all observed atmospheric depths \bs\
changes by less than about $0.5$~G per \ms\ of corresponding velocity
oscillations. Hence, possible magnetic variability is constrained to be below
1\% of the field strength.

We note that \equ\ is distinguished among the roAp stars by its strong
magnetic field, sharp lines and the exceptionally high amplitude of pulsational
line profile  changes. Yet this star defies any attempts to detect magnetic
variability with  pulsation phase. This suggests that rapid magnetic modulation
may be even more  difficult to detect in other roAp pulsators. The outcome of
our monitoring of \bz\ and \bs\ in \equ\ demonstrates that very accurate
measurements of the Zeeman resolved lines can yield more precise magnetic
time-series compared with the polarimetric \bz\ observations. Therefore, the
most promising direction for future attempts to detect magnetic oscillations in
\equ\ (which are in any case unlikely to exceed a few tens of gauss) would be to
observe those \ion{Nd}{ii}, \nd\ and \pr\ lines which are
characterized by large pulsational RV shifts and at the same time show
Zeeman resolved profiles. Unfortunately, the extra broadening of the
pulsating lines (see Kochukhov \& Ryabchikova \scite{KR01a}) strongly smears
observed Zeeman structure and makes proposed study of \bs\
oscillations extremely difficult.

\section*{Acknowledgments}
This paper is based on observations obtained at the European Southern Observatory (La Silla, Chile)
and at the Canada-France-Hawaii Telescope.
We acknowledge support by the Lise Meitner fellowship to OK (FWF project M757-N02), 
by the FWF project {\it P 14984}, by the Natural Sciences and Engineering Research Council of Canada,
by the Russian Federal program `Astronomy' (part 1102) and by RFBR (grant 04-02-16788).


\begin{thebibliography}{}

\bibitem[1998]{BBV98}
 Baldry I. K., Bedding T. R., Viskum M., Kjeldsen H., Frandsen S., 1998, MNRAS, 295, 33
\bibitem[2000]{BB00}
 Baldry I. K., Bedding T. R., 2000, MNRAS, 318, 341
\bibitem[2004]{HKB04}
 Hubrig S., Kurtz D. W., Bagnulo S., Szeifert T., Schoeller M., Mathys G., 
 Dziembowski W. A., 2004, A\&A, 415, 661
\bibitem[1998]{KH98}
 Kanaan A., Hatzes A. P., 1998, ApJ, 503, 848
\bibitem[2001a]{KR01a} 
 Kochukhov O., Ryabchikova T., 2001a, A\&A, 374, 615
\bibitem[2001b]{KR01b} 
 Kochukhov O., Ryabchikova T., 2001b, A\&A, 377, L22
\bibitem[2004]{KRP04}
 Kochukhov O., Ryabchikova T., Piskunov N., 2004, A\&A, 415, L13
\bibitem[1999]{KPR99}
 Kupka F., Piskunov N., Ryabchikova T. A., Stempels H. C., Weiss W. W.,
 1999, A\&AS, 138, 119 
\bibitem[2003]{LK03}
 Leone F., Kurtz D. W., 2003, A\&A, 407, L67  
\bibitem[1996]{MART96}
 Martinez P. et al., 1996, MNRAS, 282, 243
\bibitem[1997]{MHL97}
 Mathys G., Hubrig S., Landstreet J. D., Lanz T., Manfroid J., 1997, A\&AS 123, 353
\bibitem[1999]{PISK99} 
 Piskunov N. E., 1999, In: 2nd Workshop on Solar Polarization, J. Stenflo and K.N. Nagendra (eds.), 
 Kluwer Academic Publishers, Dodrecht, 515
\bibitem[2002]{RPK02}
 Ryabchikova T., Piskunov N., Kochukhov O., Tsymbal V., Mittermayer P.,
 Weiss W. W., 2002, A\&A, 384, 545
\bibitem[2003]{SMB03}
 Savanov I., Musaev F. A., Bondar A. V., 2003, IBVS, 5468   
\bibitem[1999]{SMR99}
 Savanov I. S., Malanushenko V. P., Ryabchikova T. A., 1999, Astron. Lett., 25, 802   
\end{thebibliography}
\end{document}